\documentclass[aps,prd, nofootinbib, preprint,11pt,a4paper]{revtex4-1}
\newcommand{\ch}[1]{#1}

\usepackage{amsmath,color}
\usepackage{amssymb}
\usepackage{epsfig}
\usepackage{graphicx}
\usepackage{hyperref}

\newcommand{\bwt}{\begin{widetext}}
\newcommand{\ewt}{\end{widetext}}
\newcommand{\be}{\begin{equation}}
\newcommand{\ee}{\end{equation}}
\newcommand{\bdm}{\begin{displaymath}}
\newcommand{\edm}{\end{displaymath}}
\newcommand{\bea}{\begin{eqnarray}}
\newcommand{\eea}{\end{eqnarray}}
\newcommand{\nn}{\nonumber}

\def\vev#1{\left\langle #1 \right\rangle}

\def\Re{\mbox{Re}\,}
\def\Tr{\mbox{Tr}\,}
\def\det{\mbox{det}\,}

\def\vare{\varepsilon}

\newcommand{\intx}
{\int_0^1\!\!{\rm d}x}

\renewcommand{\div}{{\rm C_{UV}}}

\begin{document}
\title{Fun with the Abelian Higgs Model}
\author{Michal Malinsk\'{y}}\email{malinsky@ipnp.troja.mff.cuni.cz}
\affiliation{AHEP Group, Instituto de F\'{\i}sica Corpuscular -- C.S.I.C./Universitat de Val\`encia, Edificio de Institutos de Paterna, Apartado 22085, E 46071 Val\`encia, Spain
\\
{\rm and}\\
Institute of Particle and Nuclear Physics, Faculty of Mathematics and Physics, Charles University in Prague,
V Hole\v{s}ovi\v{c}k\'{a}ch 2, 180 00 Praha 8, Czech Republic}
\begin{abstract}
In calculations of the elementary scalar spectra of spontaneously broken gauge theories there is a number of subtleties which, though often unnecessary to deal with in the order-of-magnitude type of calculations, have to be taken into account if fully consistent results are sought for. Within the ``canonical'' effective-potential approach these are, for instance: the need to handle infinite series of nested commutators of derivatives of  field-dependent mass matrices, the need to cope with spurious IR divergences emerging in the consistent  leading-order approximation and, in particular, the need to account for the fine interplay between the  renormalization effects in the one- and  two-point Green's functions which, indeed, is essential for the proper stable vacuum identification and, thus, for the correct interpretation of the results. 

In this note we illustrate some of these issues in the realm of the minimal abelian Higgs model and two of its simplest extensions including extra heavy scalars in the spectrum in attempt to exemplify the key aspects of the usual ``hierarchy problem'' lore in a very specific and simple setting.    
We emphasise that, regardless of the omnipresent polynomial cut-off dependence in the one-loop corrections to the scalar two-point function, the physical Higgs boson mass is always governed by the associated symmetry-breaking VEV and, as such, it is generally as UV-robust as all other VEV-driven masses in the theory.   
\end{abstract}
\maketitle
\tableofcontents \newpage
\section{Introduction}\label{sect:intro}
In the last three decades the progress in  high energy physics has been strongly influenced by considerations related to the so called ``hierarchy problem'' 
having to do with an apparently unnatural coexistence of vastly different energy scales in \ch{various beyond-Standard-Model constructions like, e.g., grand unification, left-right symmetric models, etc}. 

On the theory side, such discussions turned out to be extremely fruitful as they triggered an enormous amount of interest in low-energy extensions of the \ch{Standard Model (SM)} such as large extra dimensions, technicolor type of theories, TeV-scale supersymmetry and many other schemes. These were either tailored from scratch to address the hierarchy puzzle by, e.g., lowering the relevant cut-off (or even by discarding elementary scalars from the physical spectra at all), or led to its considerable alleviation by, e.g., reducing the sensitivity of the low-energy physics to the potentially large heavy-sector contributions by ensuring their (almost) exact cancellation due to an extra symmetry at play. 

Unfortunately, the hierarchy-based arguments in favour of any of these thrilling subjects are very often given in an overly rudimentary and handwaving form, usually along the lines like: ``Unlike for the fermions, the loop corrections to the light scalar masses due to super-heavy extra fields in loops yield terms which are quadratic in the heavy scale. So there is a need to re-adjust order by order the relevant bare scalar masses to an enormous degree in order to \ch{keep them} at the desired light scale; this looks unnatural and, thus, we have a problem unless there is a deeper reason why such loops are cancelled/softened/absent.''\footnote{For instance, supersymmetry achieves this by tightly relating scalar and fermionic loops which, due to the extra minus sign associated to the latter cancel each other up to the effects of the order of the soft SUSY breaking scale.}    
 
This, on one hand, makes the motivation for many popular SM extensions very short and hence(?) compelling; on the other hand, though, it tends to obscure a lot of important conceptual details which, however, should be clarified to the last bit in order to have a solid foundation for further speculative (and sometimes even quite baroque) constructions.  

The most important of these, in my opinion, is the fact that the SM Higgs \ch{field $H$} is not just {\it a} generic scalar field but it has several features which make it slightly less straightforward to adopt the simple argument above for the very Standard Model. First, the \ch{scalar doublet $\Phi$ in which $H$ resides} is charged under the $SU(2)_{L}\otimes U(1)_{Y}$ gauge symmetry and, as such, it can not couple directly to superheavy fermions (which must be vector-like from the SM point of view); hence, the relevant ``dangerous'' graphs emerge only at a higher-loop level, see e.g.~\cite{Martin:1997ab}. Nevertheless, heavy scalars $\Psi$ can always couple directly to $\Phi$ through the ``Higgs portal'' type of a renormalizable interaction $\Phi^{\dagger}\Phi\Psi^{\dagger}\Psi$ and, thus, the heavy tadpoles quadratic in $M_{\Psi}$ are always present.

More importantly, the radiative corrections due to the heavy fields coupled to the Higgs type of a scalar -- besides affecting the overall shape of the scalar potential -- shift also the position of the asymmetric vacuum of the theory (or, equivalently, destabilise the tree-level vacuum configuration; this exhibits itself by a non-zero value of the renormalized one-point Green's function at that point). Hence, the renormalized Higgs mass which is, by definition, related to the shape of the renormalized scalar potential in the neighbourhood of its true vacuum, must be evaluated with great care in the usual perturbation theory as there is more than just the notorious quadratic contributions to the relevant two-point function to be taken into account. 

\ch{In more technical terms, the physical mass of a generic scalar field $\phi$ in a spontaneously broken theory is given by the roots of the relevant renormalized two-point inverse propagator
\be
\Gamma_{\phi\phi}^{(2)}(p^{2})\equiv p^{2}-m_{\phi}^{2}-\Sigma_{\phi\phi}(p^{2})=0\,,\label{twopointfunction}
\ee
where $m^{2}$ corresponds to the (scheme-dependent) renormalized mass parameter and $\Sigma(p^{2})$ is the properly renormalized loop factor\footnote{\ch{The specific shape of $\Sigma(p^{2})$ in terms of the sum of the regularized loop diagrams and the appropriately fixed counterterms is, of course, renormalization-scheme dependent; however, in any given scheme the physical mass of the relevant scalar still obeys Eq.~(\ref{twopointfunction}) because in such a case also the interpretation of the bare parameters changes accordingly.}},
if and only if one sits in the proper renormalized vacuum of the theory, i.e., iff
\be
\Gamma_{\phi}^{(1)}=0\, \text { for all }\phi.\label{onepointfunction}
\ee
The last condition is, however, trivially fulfilled for all scalars that are charged with respect to the desired residual symmetry, i.e., those that can not get a vacuum expectation value (VEV); for such fields Eq.~(\ref{twopointfunction}) contains the full information. On the other hand, for the scalars of the Higgs type, i.e., for those that {\it can} acquire a non-zero VEV, Eq.~(\ref{onepointfunction}) represents a non-trivial extra constraint that has to be implemented into (\ref{twopointfunction}) before its roots are determined. 
} 

However, as important as it is, this subtlety is almost never addressed in courses or introductory lectures; this, in turn, generates a bias against elementary scalar fields without making any distinction between those having nothing to do with spontaneous symmetry breaking and the Higgs-like degrees of freedom. 

\vskip 3mm 
In this note we attempt to provide an elementary review of the Higgs mass renormalization in  the simplest gauge theory that can accommodate perturbative spontaneous symmetry breaking, the minimal abelian Higgs model. In particular, we would like to demonstrate in all detail that the large quadratic contributions due to the superheavy fields (with masses $\sim M$)  can be entirely subsumed into the renormalization of the relevant VEV so that there is no other explicit $M^{2}$-dependence left in the {\it physical}\, Higgs mass.

\vskip 2mm 

Generalising this to the SM situation one may thus conclude that, at variance with what one can hear frequently, the physical mass of the SM Higgs is not any more sensitive to the UV physics than that of any other SM field.

\vskip 5mm
In doing so, it will be very convenient to  invoke the effective potential techniques as an ideal bookkeeping tool that saves one from most of the explicit diagrammatics hassle. However, in this approach (and, in particular, in spontaneously broken gauge theories) there are other subtleties one should be aware of, e.g., the need to handle infinite series of nested commutators of derivatives of  field-dependent mass matrices, the need to cope with spurious IR divergences emerging in the self-consistent  leading-order approximation and also the need to account for the finite shifts from the zero momentum regime (the effective potential as the zero-derivative term in the coordinate expansion of the effective action corresponds to) to the mass shell. Since this, I believe, is an interesting stuff per se we shall spend some time on such technicalities too; hopefully, this will make these notes  more self-contained and potentially interesting for a wider readership. 
\vskip 5mm
\paragraph*{Outline:}
After few formal prerequisites related to practical aspects of the effective potential  approach, cf. Section~\ref{sect:prerequisites}, in Section~\ref{sect:minimalAHM} we shall give a short review of some of the salient features of the abelian Higgs model: 
\begin{itemize}
\item
We shall show explicitly how important it is to keep in mind that the field-dependent mass form is actually a matrix that does not need to commute with its derivatives and how this is reflected by the presence of infinite-series terms in the second derivatives of the one-loop effective potential. This, as we shall see, is indeed crucial to keep the Goldstone modes formally massless at the one-loop level.
\item We shall discuss in brief how to deal with apparent IR divergences due to the presence of massless modes (recall that the Goldstone-boson propagator in the Landau gauge is IR singular) emerging in a simple-minded (though perturbatively consistent) leading-order approximation or even in the full-fledged one-loop calculation within a particularly contrived renormalization scheme. 
\item We shall also try to make it clear that, in spite of the explicit scale factor present in various versions of the renormalized Higgs mass formula, the physical Higgs mass is unique and scale independent. 
\end{itemize}

Then, in Sect.~\ref{sect:AHMplusscalar}, we shall repeat the one-loop spectrum calculation in a non-minimal model featuring an extra scalar field $\Psi$ equipped with a large gauge-singlet mass term $M$. It will become very clear that the physical mass of the Higgs boson is prone to large radiative corrections quadratic in $M$ only through \ch{the associated symmetry breaking} VEV \ch{while all its explicit $M^{2}$-dependence is only logarithmic}. In this respect, the Higgs bosons are as UV robust as all the other fields whose masses are governed by the same VEV (\ch{such as,} e.g., the $U(1)$ gauge boson in the abelian Higgs model under consideration). 

Finally, in Sect.~\ref{sect:AHMplustwoscalars}, we shall argue that this is a  special feature of the Higgs bosons (i.e., \ch{the dynamical components of the scalar fields} triggering a spontaneous symmetry breakdown) as it does not apply to other generic ``accidentally'' light scalars -- redoing the same calculation yet again with an accidentally light extra scalar singlet \ch{(with respect to the unbroken gauge symmetry)} we shall see explicitly that the loop corrections drag its mass to the heavy scale as expected while leaving the Higgs mass essentially intact.   

For the sake of completeness a set of appendices including  computational details is attached. 
\section{Few formal prerequisites\label{sect:prerequisites}}
Assuming just scalar and gauge degrees of freedom at play the one-loop (Coleman-Weinberg) effective potential~\cite{Coleman:1973jx} reads $V_{\rm eff}=V+\Delta V$ where $V$ denotes the tree-level scalar potential,
\begin{equation}
\label{DeltaV}
\Delta V(\phi,\mu) =
\frac{1}{64\pi^2}
\Tr\left[{M}_{S}^4(\phi)\left(\log\!\frac{{M}_{S}^2(\phi)}{\mu^2}
-\frac{3}{2}\right)+3{M}_{G}^4(\phi)\left(\log\!\frac{{M}_{G}^2(\phi)}{\mu^2}
-\frac{5}{6}\right)\right] 
\end{equation}
is the relevant one-loop correction \ch{(in the dimensional regularisation within the modified minimal subtraction ($\overline{\rm MS}$) scheme)} and ${M}_{S}$ and ${M}_{G}$ stand for the matrices of the second derivatives of  $V$ and the field-dependent gauge masses, respectively. Let us also note that $\mu$ is the renormalization scale and all parameters entering (\ref{DeltaV}) are understood to be the running parameters depending implicitly on $\mu$ with the additive factors in the round brackets ensuring their proper interpretation \ch{in the given renormalization scheme}. 

In what follows we shall focus predominantly on the scalar sector of the model, i.e., we shall mostly ignore the second (gauge) term in the square bracket of (\ref{DeltaV}), thus avoiding all the subtleties related to the gauge dependence of the effective potential etc. An interested reader can find a thorough discussion of these matters, e.g., in \cite{Dolan:1974gu,Nielsen:1975fs,Fukuda:1975di} and references therein.  We shall also entirely ignore other effects due to, for instance, finite temperature~\cite{Kirzhnits:1972iw,Kirzhnits:1972ut} and/or finite density~\cite{Kapusta:1981aa,Haber:1981ts}. 

In theories with more than a single scalar \ch{field (i.e., in every perturbative implementation of the Higgs mechanism where the longitudinal components of the massive vector bosons are supplied by the Goldstones)}  ${M}_{S}$ must be treated as a matrix and the logs are then defined through the relevant series.  Since the pre-log matrices in (\ref{DeltaV}) are identical to those in the log arguments, their (trivial) commutativity properties ensure that the first derivative of $\Delta V$ can be written in a simple form:
\begin{equation}
\label{der1deltaV}
\frac{\partial\Delta V}{\partial\phi}=
\frac{1}{64\pi^2}\Tr\left[\left\{\frac{\partial{M}_{S}^2}{\partial\phi},
  {M}_{S}^2\right\}  
\left(\log\!\frac{{M}_{S}^2}{\mu^2}-\frac{3}{2}\right)
+ {M}_{S}^2\frac{\partial{M}_{S}^2}{\partial\phi}\right]+{\rm gauge\;part} \,,
\end{equation}
Unfortunately, this is no longer the case for the higher derivatives as there is no guarantee that ${M}_{S}$ in the logs in formula~(\ref{der1deltaV}) commutes with its \ch{derivatives} in the pre-log factor. This, however, complicates the evaluation of the second derivatives, giving rise to an infinite series of terms including nested commutators of these structures:
\begin{eqnarray}
&& \frac{\partial^2\Delta V}{\partial\phi_a\partial\phi_b}=\nonumber \\
&& \frac{1}{64\pi^2} \Tr \Biggl\{\left(\left\{\frac{\partial^{2} {M}_{S}^2}{\partial\phi_a\partial\phi_b}, {M}_{S}^2\right\}
+ \left\{ \frac{\partial {M}_{S}^2}{\partial\phi_a}, \frac{\partial {M}_{S}^2}{\partial\phi_b}\right\}\right)\left(\log\!\frac{{M}_{S}^2}{\mu^2}-\frac{3}{2}\right)+\frac{\partial {M}_{S}^2}{\partial\phi_a}\frac{\partial {M}_{S}^2}{\partial\phi_b}
+{M}_{S}^2\frac{\partial^{2} {M}_{S}^2}{\partial\phi_a\partial\phi_b}\nonumber \\
&&\quad+ \sum_{m=1}^{\infty}(-1)^{m+1}\frac{1}{m}\sum_{k=1}^{m}{m\choose k}\left\{ \frac{{M}_{S}^{2}}{\mu^{2}}, \frac{\partial {M}_{S}^2}{\partial\phi_a}\right\}  
\left[\frac{{M}_{S}^{2}}{\mu^{2}},..\left[\frac{{M}_{S}^{2}}{\mu^{2}}, \frac{\partial {M}_{S}^2}{\partial\phi_b}\right]..\right]\left(\frac{{M}_{S}^{2}}{\mu^{2}}-1\right)^{m-k}
\Biggr\}+{\rm gauge\;part}\,,\nonumber\\
\label{der2deltaV}
\end{eqnarray}
where, for each $k\geq 1$, the commutator in the last term is taken $k-1$ times. Let us note that the RHS of formula (\ref{der2deltaV}) can be shown to be symmetric under $a\leftrightarrow b$, as it should be.
Though more difficult to handle in practice the infinite series of the nested commutators plays a central role in rendering our later calculations self-consistent. 

\section{The minimal abelian Higgs model\label{sect:minimalAHM}}
\subsection{The classical lagrangian and the tree-level vacuum}
In what follows we shall first consider the classical abelian Higgs model lagrangian
\be
{\cal L}=-\frac{1}{4}F_{\mu\nu}F^{\mu\nu}+(D_{\mu}\Phi)^{\dagger}D^{\mu}\Phi-V(\Phi^{\dagger}\Phi)\,,
\ee
with a renormalizable scalar potential in the form
\be
V(\Phi^{\dagger}\Phi)=-m^{2}\Phi^{\dagger}\Phi+\lambda (\Phi^{\dagger}\Phi)^{2}\,.
\ee
Writing $\Phi=R e^{i\alpha}$ and assuming $R\geq 0$ and $m^{2}>0$ the rewritten scalar potential 
\be
V(R,\alpha)=-\frac{1}{2}m^{2}R^{2}+\lambda R^{4}
\ee
is minimized for $R=m/\sqrt{\lambda}$ and an arbitrary $\alpha$. 
Choosing without loss of generality $\left\langle\Phi\right\rangle= v\in \mathbb{R}$ the physical Higgs (H) and the Goldstone (G) modes are readily identified as  
\be
\Phi=\frac{1}{\sqrt{2}}(H+v+iG)
\ee
and, up to an irrelevant constant term, the scalar potential reads  in the broken phase
\bea\label{treelagrangianasymmetric}
V(H,G) & =& (-m^2+3\lambda v^{2})\frac{H^2 }{2} + (-m^2+\lambda v^{2})\frac{G^2 }{2}\nn\\
&+&H v(-m^2 +\lambda v^{2})+\lambda v (H^3+HG^{2})+\frac{\lambda }{4}(H^4+2H^{2} G^{2}+G^{4})
\eea
which, in the tree-level minimum $v=m/\sqrt{\lambda}$, receives the familiar form
\be
V(H,G)=\frac{1}{2}(2\lambda v^{2})H^{2}+\lambda vH^{3}+\lambda vHG^{2}+\frac{1}{4}\lambda H^{4}+\frac{1}{4}\lambda G^{4}+\frac{1}{2}\lambda H^{2}G^{2}\,.
\ee
Hence, at the tree level, the physical Higgs mass is 
\be
\label{treemH2}
m_{H}^{2}=2\lambda v^{2}\,.
\ee
\subsection{The one-loop vacuum and the second derivatives of the effective potential\label{prereqs:singleHiggs}}
The central object of interest in the calculation of the  one-loop Higgs mass is the tree-level field-dependent scalar mass matrix  (defined as a matrix of second derivatives of $V$ with respect to all scalar degrees of freedom) and its first and second derivatives evaluated at the minimum of the one-loop effective potential, cf. equation (\ref{der2deltaV}). Restoring the general coordinates in the scalar sector
, $\Phi=\phi_{R}+i \phi_{I}$, one has 
\be
V(\phi_{I},\phi_{R})=-\frac{1}{2}m^{2}({\phi_{R}^{2}+\phi_{I}^{2}})+\frac{1}{4}\lambda({\phi_{R}^{2}+\phi_{I}^{2}})^{2}\,,
\ee
and the field-dependent mass matrix in the $\{\phi_{R},\phi_{I}\}$ basis reads
\be\label{MSsquaredminimal}
{M}_{S}^{2}=
\left(\begin{array}{cc}
-m^{2}+3\lambda \phi_{R}^{2}+\lambda \phi_{I}^{2} & 2 \lambda \phi_{R} \phi_{I}  \\
2 \lambda \phi_{R} \phi_{I}  & -m^{2}+3\lambda \phi_{I}^{2}+\lambda \phi_{R}^{2}
\end{array}\right)\,.
\ee 
In the same basis,
\be
\frac{\partial{M}_{S}^{2}}{\partial\phi_{R}}=
\left(\begin{array}{cc}
6\lambda \phi_{R} & 2 \lambda  \phi_{I}  \\
2 \lambda \phi_{I}  & 2\lambda \phi_{R}
\end{array}\right)\,,\qquad
\frac{\partial{M}_{S}^{2}}{\partial\phi_{I}}=
\left(\begin{array}{cc}
2\lambda \phi_{I} & 2 \lambda  \phi_{R}  \\
2 \lambda \phi_{R}  & 6\lambda \phi_{I}
\end{array}\right)\,,
\ee 
and
\be\label{MSsquaredminimal2derivatives}
\frac{\partial^{2}{M}_{S}^{2}}{\partial\phi_{R}^{2}}=
\left(\begin{array}{cc}
6\lambda  & 0  \\
0  & 2\lambda 
\end{array}\right)\,,\qquad
\frac{\partial^{2}{M}_{S}^{2}}{\partial\phi_{I}^{2}}=
\left(\begin{array}{cc}
2\lambda & 0  \\
0 & 6\lambda
\end{array}\right)\,,\qquad
\frac{\partial^{2}{M}_{S}^{2}}{\partial\phi_{R}\partial\phi_{I}}=
\left(\begin{array}{cc}
0 & 2\lambda \\
2\lambda & 0
\end{array}\right)\,.
\ee 
These structures, evaluated at the assumed\footnote{\ch{Needless to say, the assumed reality of $\vev\Phi$ does not lead to any loss of generality.}} real asymmetric vacuum $\left\langle \Re\Phi\right\rangle=v$ (i.e., $\left\langle\phi_{R}\right\rangle=v$, $\left\langle\phi_{I}\right\rangle=0$), provide all the ingredients needed to construct the first and second derivatives in eqs.~(\ref{der1deltaV}) and (\ref{der2deltaV}) of the (scalar part of the) one-loop effective potential of our interest.
\paragraph*{First derivatives:}
Using Eq.~(\ref{der1deltaV}) the first derivatives of the one-loop effective potential evaluated at the asymmetric vacuum are zero if and only if  (as before, we consider only the scalar part of Eq.~(\ref{DeltaV}))
\bea\label{stationarycondition}
& & -m^{2}+\lambda v^{2}+\frac{\lambda}{16\pi^{2}}\left[4m^{2}-10\lambda v^{2}-m^{2}\log\!\left(\frac{-m^{2}+\lambda v^{2}}{\mu^{2}}\right)-3m^{2}\log\!\left(\frac{-m^{2}+3\lambda v^{2}}{\mu^{2}}\right)\right.\\
& &\left.\hspace{5.5cm}+\lambda v^{2}\log\!\left(\frac{-m^{2}+\lambda v^{2}}{\mu^{2}}\right)+9\lambda v^{2}\log\!\left(\frac{-m^{2}+3\lambda v^{2}}{\mu^{2}}\right)\right]=0\nn\,,
\eea
which constitutes a one-loop stationarity condition for the VEV $v$ in terms of the scalar potential parameters.
\paragraph*{Second derivatives:}
Since ${M}_{S}^{2}$ and  $\partial{M}_{S}^{2}/\partial \phi_{I}$ evaluated at the vacuum do not commute, one expects that the second derivatives of the scalar part ($\Delta V_{S}$) of the one-loop contribution  to the effective potential (\ref{der2deltaV}), namely, the second derivative of $\Delta V_{S}$ with respect to $\phi_{I}$ and the mixed one, will require a careful summation of the relevant infinite series. On the other hand, since ${M}_{S}^{2}$ and  $\partial{M}_{S}^{2}/\partial \phi_{R}$ evaluated at the asymmetric vacuum do commute, the second derivative of $\Delta V_{S}$ with respect to $\phi_{R}$ should be much simpler to deal with.

Using the methods described in Appendix~\ref{sect:handlingtheseries} the relevant series yield the following sums (in an obvious notation):
\bea
S_{RR}&=&\left(\begin{array}{cc}
72 \lambda^{2}v^{2} & 0 \\
0 & 8\lambda^{2}v^{2} 
\end{array}\right)\,,\\
S_{II}&=&4 \lambda(m^{2}-2\lambda v^{2})\log\!\left(\frac{-m^{2}+\lambda v^{2}}{-m^{2}+3\lambda v^{2}}\right)\left(\begin{array}{cc}
1 & 0 \label{SGG}\\
0 &1
\end{array}\right)\,,
\eea
where we displayed only those factors that can contribute to (\ref{der2deltaV}), i.e., those that are not traceless. 

Given these prerequisites, one can finally write down the full matrix of the second derivatives of the (scalar part of the) effective potential:
\bea
\left\langle\frac{\partial^{2}V_{\rm eff}}{\partial \phi_{R}^{2}}\right\rangle
&=& -m^{2}+3\lambda v^{2}+\frac{\lambda}{16\pi^{2}}\left[4m^{2}-10\lambda v^{2}-m^{2}\log\!\left(\frac{-m^{2}+\lambda v^{2}}{\mu^{2}}\right)\!-3m^{2}\log\!\left(\frac{-m^{2}+3\lambda v^{2}}{\mu^{2}}\right)\right.\nn\\
& &\left.\hspace{5cm}+3\lambda v^{2}\log\!\left(\frac{-m^{2}+\lambda v^{2}}{\mu^{2}}\right)+27\lambda v^{2}\log\!\left(\frac{-m^{2}+3\lambda v^{2}}{\mu^{2}}\right)\right],\nn\\
\\
\left\langle\frac{\partial^{2}V_{\rm eff}}{\partial \phi_{I}^{2}}\right\rangle
&=& -m^{2}+\lambda v^{2}+\frac{\lambda}{16\pi^{2}}\left[4m^{2}-10\lambda v^{2}-m^{2}\log\!\left(\frac{-m^{2}+\lambda v^{2}}{\mu^{2}}\right)-3m^{2}\log\!\left(\frac{-m^{2}+3\lambda v^{2}}{\mu^{2}}\right)\right.\nn\\
& &\left.\hspace{5.3cm}+\lambda v^{2}\log\!\left(\frac{-m^{2}+\lambda v^{2}}{\mu^{2}}\right)+9\lambda v^{2}\log\!\left(\frac{-m^{2}+3\lambda v^{2}}{\mu^{2}}\right)\right],\nn\\
\label{secondderivatives}
\eea
where only the non-zero elements were displayed.
Remarkably, in the exact one-loop minimum given by the stationary condition (\ref{stationarycondition})  the ``Goldstone part'' $\left\langle{\partial^{2}V_{\rm eff}}/{\partial \phi_{I}^{2}}\right\rangle$ vanishes, as required by the overall  consistency. Note that the infinite series in (\ref{der2deltaV}) and, in particular, its sum (\ref{SGG}) play a central role in this calculation -- without it one could never get an exactly massless Goldstone at the one-loop level.

\subsection{The one-loop Higgs mass\label{oneloopHiggssimplest}}
Finally, turning to the Higgs mass (now it is already obvious that $H=\phi_{R}$ up to the VEV) one has to evaluate the first formula in~(\ref{secondderivatives}) at the one-loop vacuum~(\ref{stationarycondition}). Although it is not possible to solve Eq.~(\ref{stationarycondition})  for $m^{2}$ in a closed form it is still very useful to subtract it from $\left\langle{\partial^{2}V_{\rm eff}}/{\partial \phi_{R}^{2}}\right\rangle$; this yields 
\be\label{Higgszeromomentummass}
\left\langle\frac{\partial^{2}V_{\rm eff}}{\partial \phi_{R}^{2}}\right\rangle
= 2\lambda v^{2}+\frac{\lambda^{2}v^{2}}{8\pi^{2}}\log\!\left(\frac{-m^{2}+\lambda v^{2}}{\mu^{2}}\right)+\frac{9\lambda^{2}v^{2}}{8\pi^{2}}\log\!\left(\frac{-m^{2}+3\lambda v^{2}}{\mu^{2}}\right)\,,
\ee
where the first term on the RHS is obviously the tree-level contribution~(\ref{treemH2}).

There are several comments worth making here:
\begin{itemize}
\item 
It is very instructive to trace back the disappearance of the terms \ch{quadratic in $m$} and also the $m^{2}$-proportional pre-factors of logs; although this can be viewed as a trivial consequence of trading $m^{2}$ for $\lambda v^{2}+\ldots$ due to~(\ref{stationarycondition}) the same, as we shall see,  happens even if there are other large singlet mass parameters at play due to, e.g., heavy extra singlet scalars, cf. Section~\ref{sect:AHMplusscalar}. 
\item
It can look like the first log in formula~(\ref{Higgszeromomentummass}) contains an IR singularity if one uses a tree-level stationarity condition $m^{2}=\lambda v^{2}$ here (which, operationally, can be justified by arguing that the error one would commit if the loop corrections in~(\ref{stationarycondition}) were neglected is of a higher order). On a little less naive basis one can, e.g.,  come up with a contrived renormalization scheme in which $m^{2}=\lambda v^{2}$ would be an exact equation for the one-loop vacuum; indeed, it is sufficient to choose $\mu$ such that~(\ref{stationarycondition}) holds, i.e.,   
$\log\!\left({2\lambda v^{2}}/{\mu^{2}}\right)=1$
which blows the first log in~(\ref{Higgszeromomentummass}) too. 
What does such an instability mean?  Note that there is nothing like this in the Goldstone sector as the formula (\ref{secondderivatives}) is functionally identical to the stationarity condition~(\ref{stationarycondition}).  

The key to this issue is the fact that the effective potential corresponds to the first term in the momentum expansion of the effective action {\it around zero momentum} and, thus,  equation~(\ref{Higgszeromomentummass}) does not correspond to the physical (i.e., pole) mass of a massive scalar. This, indeed, is obtained by shifting the zero-momentum mass~(\ref{Higgszeromomentummass}) to the momentum squared identical to the pole mass squared, i.e., to solving the secular equation of the relevant eigenvalue problem
\be\label{polemass}
\det[{\cal M}^{2}-p^{2}+\Sigma(p^{2})-\Sigma(0)]=0\,,
\ee
\ch{(see Appendix~\ref{AppFiniteShifts})} where ${\cal M}^{2}$ is the zero-momentum mass matrix given by the second derivatives of the one-loop effective potential~(\ref{secondderivatives}) and $\Sigma$ is the matrix of the  scalar self-energies. In this respect, it is clear why such an instability can not occur in the Goldstone sector - for zero $p^{2}$-eigenvalues the self-energies drop from Eq.~(\ref{polemass}) and, thus, there is no extra contribution to save the day. 

The (scalar part of the) self-energy matrix $\Sigma$ is calculated in Appendix~\ref{sect:selfenergies} 
and it is clear that, indeed, the momentum dependent piece of $\Sigma_{HH}$ has an IR singularity at $p^{2}=0$, $m^{2}=2\lambda v^{2}$ which matches exactly the singularity in the first log of Eq.~(\ref{Higgszeromomentummass}). Thus, there is no IR singularity in the {\it physical} mass of the Higgs boson determined from (\ref{polemass}) at the one-loop level.

\item
It is not only the first log in~(\ref{Higgszeromomentummass}) that can be expected on the consistency grounds to appear in the second derivative of the one-loop effective potential; also the second log can be guessed without entering the tedium of an explicit effective potential calculation. The argument goes as follows: Even if the apparent IR singularity in~(\ref{Higgszeromomentummass}) is tamed by the $\Sigma_{HH}(p^{2})-\Sigma_{HH}(0)$ difference, this extra shift does not affect the explicit $\mu$-dependence of the RHS of Eq.~(\ref{Higgszeromomentummass}). However, unlike the running mass, the pole mass is a fixed number which should not depend on the renormalization scale so for a full conceptual consistency the explicit $\mu$-dependence in the one-loop formula for the physical Higgs mass should be compensated by the implicit $\mu$-dependence of the running quantities in there. Barring higher order (two-loop) terms the only piece whose implicit $\mu$-dependence can compete with the explicit $-5\lambda^{2}v^{2}/4\pi^{2}\log\!(\mu^{2})\equiv X(\mu)$ term on the RHS of Eq.~(\ref{Higgszeromomentummass}) is the tree-level Higgs mass $2\lambda v^{2}$. Since the (scalar part of the) one-loop anomalous dimension of the Higgs field in the minimal abelian Higgs model is zero, cf. Appendix~\ref{sect:anomalousdimension}, the culprit should be the quartic coupling $\lambda$. Indeed, using the scalar part of the relevant beta-function calculated in Appendix~\ref{sect:beta} one can see that a shift in the renormalization scale $\mu^{2}\to \tilde\mu^{2}$ inflicts a shift 
\be\label{treelevelshift}
2\lambda(\mu) v^{2}\to 2\lambda(\tilde\mu)v^{2}=2\lambda(\mu) v^{2}+\frac{5\lambda^{2}v^{2}}{4\pi^{2}}\log\!\left(\frac{\tilde\mu^{2}}{\mu^{2}}\right)
\ee
in the tree-level contribution while the one-loop part in~(\ref{Higgszeromomentummass}) gets shifted as 
\be
X(\mu)\to X(\tilde\mu)=X(\mu)-\frac{5\lambda^{2}v^{2}}{4\pi^{2}}\log\!\left(\frac{\tilde\mu^{2}}{\mu^{2}}\right)\,.
\ee
Consequently, the total $2\lambda(\mu) v^{2}+X(\mu)$ remains $\mu$-independent at one loop. 
\end{itemize}
Reversing the logic, i.e., demanding that there is no ``spurious'' IR divergence in the one-loop Higgs formula and that the Higgs pole mass is renormalization-scale independent one can fully reconstruct the formula (\ref{Higgszeromomentummass}) without even writing down the one loop effective potential, let alone dealing with the series of nested commutators!
To conclude, the result~(\ref{Higgszeromomentummass}) is fully justified (including several conceptual details) and we are ready to generalize it for the case of our main interest. 

Yet another comment is in place though: It is, of course, not an accident that the \ch{finite} part of \ch{$\Sigma_{HH}(p^{2})$ at $p^{2}=0$, see formula~(\ref{SigmaExplicit}),} corresponds exactly to the one-loop part of Eq.~(\ref{Higgszeromomentummass})\ch{; this is even to be expected because -- after the finite shift\footnote{\ch{Let us reiterate that the effective potential entails  only the zeroth-order term in the momentum expansion of the effective action.}} given by the last two terms in formula~(\ref{polemass}) -- one should get the notorious expression
\be\label{naiveformula}
m_{H}^{2}=2\lambda v^{2}+\Sigma_{HH}(p^{2}=2\lambda v^{2})|_{\text{finite part}}\,,
\ee 
which is the correct formula for the root of the inverse propagator up to higher-order corrections.

Nevertheless, one should not get the false impression that all the effective potential tedium is useless because the physical Higgs mass could have been calculated from scratch by just summing up the tree-level piece $2\lambda v^{2}$ with $\Sigma_{HH}(p^{2})$ evaluated at $p^{2}=2\lambda v^{2}$ with the UV-divergent part subtracted\footnote{\ch{Indeed, this is the expression that is usually written down in  support of the arguments about the explicit quadratic Higgs mass sensitivity to heavy degrees of freedom.}}. 
The point here is that this simple procedure does not in any way reflect the quantum effects in the vacuum of the theory as it assumes implicitly that $m^{2}=\lambda v^{2}$ is an excellent approximation of the shape of the vacuum manifold even at the one-loop level. This, however, can be far from true if there are heavy degrees of freedom coupled to the Higgs; the associated tadpoles may easily induce large shifts in the relevant one-point function (\ref{onepointfunction}) quadratic in their mass. Similarly, also $\Sigma_{HH}(p^{2})|_{\text{finite part}}$ can contain large quadratic contributions from the heavy seagul type of diagrams which -- if evaluated in the na\"\i ve minimum -- would, indeed, yield an explicit large-mass-squared dependence. Hence, the interplay between the one- and two-point functions is essential for the proper  physical Higgs mass identification at the quantum level. 
}

It should also be clear that all the simple links between the effective potential and the classical diagrammatic calculation that we drew for the abelian Higgs model would be much more difficult to follow in more elaborate scenarios. For that sake it is sufficient to compare the elegance of the former (where, in fact, the main diagrammatic tedium is contained in calculating few derivatives of the prefabricated formula~(\ref{DeltaV})) with, e.g., the full-fledged calculation of all the diagrams in realistic theories such as the Standard Model~\cite{Ford:1992pn} and/or the myriads of its potentially realistic extensions\footnote{For a recent effective potential analysis of the minimal $SO(10)$ grand unified theory see, for instance,~\cite{Bertolini:2009es}.}.
Moreover, the advantage of the effective potential language becomes further pronounced when the quantum structure of not only the one- and two-point Green's functions is at stakes (i.e., when it comes to, e.g., vacuum stability,  thermal effects etc.), let alone the deep conceptual questions as, for instance, the renormalizability of \ch{the spontaneously broken gauge theories}~\cite{0521318270}.
\section{The abelian Higgs model with an extra singlet scalar\label{sect:AHMplusscalar}}
\subsection{The Higgs Anti-discrimination Act}
In this section we shall recalculate the one-loop mass of the abelian Higgs boson in the presence of an extra singlet scalar $\Psi$ whose mass is governed by a large gauge-singlet\footnote{Of course, the gauge-singlet nature of the scalar mass here is implied and does not need to be emphasized; what we mean  is namely that $\Psi$ has nothing to do with the spontaneous breaking of the gauge symmetry.} mass term $\tfrac{1}{2}M^{2}\Psi^{2}$. Our main motivation is to exemplify that even with an extra (super-heavy) field coupled to the Higgs boson the physical mass of the latter is still given by a formula similar to~(\ref{Higgszeromomentummass}). This should make it clear that the large corrections quadratic in $M$ (or, equivalently, in the cut-off scale $\Lambda$ where a new physics is supposed to kick in) are all subsumed into a shift of the \ch{relevant} VEV and, besides that, there are no further $M^{2}$-like contributions to the {\it physical} Higgs mass. Put another way, even at the loop level the Higgs mass is governed by \ch{the VEV of its host scalar field} and all the explicit $M$ dependence turns out to be only logarithmic. Hence, as we anticipated, {\it the mass of the Higgs in the abelian Higgs model is as UV robust as that of the associated gauge field}. \ch{Therefore}, the Higgs should not be accused from causing trouble with naturalness any more than other fields with VEV-driven masses. 
\subsection{The lagrangian}
Assuming for simplicity that the heavy neutral singlet $
\Psi$ is odd\footnote{This is an entirely technical assumption which simplifies the structure of the scalar potential. Relaxing this does not change the qualitative features of the theory in any substantial way.} under a $Z_{2}$ symmetry $\Psi \to -\Psi$ the relevant lagrangian can be written in the form
\be
{\cal L}=-\frac{1}{4}F_{\mu\nu}F^{\mu\nu}+(D_{\mu}\Phi)^{\dagger}D^{\mu}\Phi+\frac{1}{2}\partial_{\mu}\Psi\partial^{\mu}\Psi-V(\Phi^{\dagger}\Phi,\Psi^{2})\,,
\ee
where the renormalizable scalar potential reads
\be
V(\Phi^{\dagger}\Phi,\Psi^{2})=-m^{2}\Phi^{\dagger}\Phi+\lambda (\Phi^{\dagger}\Phi)^{2}+\frac{1}{2}M^{2}\Psi^{2}+\rho \Psi^{4}+\kappa \Phi^{\dagger}\Phi\Psi^{2}\,,
\ee
which, in components, looks
\be\label{VWithPsiInComponents}
V(\phi_{I},\phi_{R})=-\frac{1}{2}m^{2}({\phi_{R}^{2}+\phi_{I}^{2}})+\frac{1}{4}\lambda({\phi_{R}^{2}+\phi_{I}^{2}})^{2}+\frac{1}{2}M^{2}\Psi^{2}+\rho \Psi^{4}+\frac{1}{2}\kappa({\phi_{R}^{2}+\phi_{I}^{2}})\Psi^{2}\,.
\ee

\subsection{The one-loop vacuum and the second derivatives of the effective potential}
Following the same procedure as in Section \ref{prereqs:singleHiggs} the  relevant analogues of eqs.~(\ref{MSsquaredminimal})-(\ref{MSsquaredminimal2derivatives}) read here (in the $\{\phi_{R},\phi_{I},\Psi\}$ basis):
\be\label{MSsquarednexttominimal}
{M}_{S}^{2}\!=\!
\left(\begin{array}{ccc}
-m^{2}+3\lambda \phi_{R}^{2}+\lambda \phi_{I}^{2}+\kappa\Psi^{2} & 2 \lambda \phi_{R} \phi_{I} & 2\kappa \phi_{R}\Psi \\
2 \lambda \phi_{R} \phi_{I}  & -m^{2}+3\lambda \phi_{I}^{2}+\lambda \phi_{R}^{2}+\kappa\Psi^{2} & 2\kappa \phi_{I}\Psi \\
2\kappa \phi_{R}\Psi & 2\kappa \phi_{I}\Psi & M^{2}+\kappa(\phi_{R}^{2}+\phi_{I}^{2})+12\rho\Psi^{2}
\end{array}\right)
\ee 
\be
\frac{\partial{M}_{S}^{2}}{\partial\phi_{R}}\!=\!
\left(\!\begin{array}{ccc}
6\lambda \phi_{R} & 2 \lambda  \phi_{I}  & 2\kappa \Psi\\
2 \lambda \phi_{I}  & 2\lambda \phi_{R} & 0\\
2\kappa \Psi & 0 & 2\kappa \phi_{R}
\end{array}\!\right),\,
\frac{\partial{M}_{S}^{2}}{\partial\phi_{I}}\!=\!
\left(\!\begin{array}{ccc}
2\lambda \phi_{I} & 2 \lambda  \phi_{R} & 0  \\
2 \lambda \phi_{R}  & 6\lambda \phi_{I} & 2\kappa \Psi \\
0 & 2\kappa \Psi & 2\kappa \phi_{I}
\end{array}\!\right),\,
\frac{\partial{M}_{S}^{2}}{\partial\Psi}\!=\!
\left(\!\begin{array}{ccc}
2\lambda \phi_{I} & 2 \lambda  \phi_{R} & 2\kappa \phi_{R}  \\
2 \lambda \phi_{R}  & 6\lambda \phi_{I} & 2\kappa \phi_{I} \\
2\kappa \phi_{R}& 2\kappa \phi_{I} & 24\rho \Psi
\end{array}\!\right)
\ee 
and
\be
\frac{\partial^{2}{M}_{S}^{2}}{\partial\phi_{R}^{2}}=
\left(\begin{array}{ccc}
6\lambda  & 0 & 0  \\
0  & 2\lambda & 0 \\
0& 0 & 2\kappa
\end{array}\right)\,,\qquad
\frac{\partial^{2}{M}_{S}^{2}}{\partial\phi_{I}^{2}}=
\left(\begin{array}{ccc}
2\lambda & 0  & 0 \\
0 & 6\lambda & 0 \\
0& 0 & 2\kappa
\end{array}\right)\,,\qquad
\frac{\partial^{2}{M}_{S}^{2}}{\partial\Psi^{2}}=
\left(\begin{array}{ccc}
2\kappa & 0 & 0  \\
0 & 2\kappa & 0 \\
0 & 0 & 24\rho 
\end{array}\right)\,,
\ee 
\be\label{MSsquarednexttominimal2derivatives}
\frac{\partial^{2}{M}_{S}^{2}}{\partial\phi_{R}\partial\phi_{I}}=
\left(\begin{array}{ccc}
0 & 2\lambda & 0  \\
2\lambda & 0 & 0 \\
0 & 0 & 0 
\end{array}\right)\,,\qquad
\frac{\partial^{2}{M}_{S}^{2}}{\partial\phi_{R}\partial\Psi}=
\left(\begin{array}{ccc}
0 & 0  & 2\kappa \\
0 & 0 & 0 \\
2\kappa& 0 & 0
\end{array}\right)\,,\qquad
\frac{\partial^{2}{M}_{S}^{2}}{\partial\phi_{I}\partial\Psi}=
\left(\begin{array}{ccc}
0 & 0 & 0  \\
0 & 0 & 2\kappa \\
0 & 2\kappa & 0 
\end{array}\right)\,.
\ee 
\paragraph*{First derivatives:}
As expected, the simple one-loop stationarity condition (\ref{stationarycondition}) receives an extra set of $M^2$-proportional terms due to the heavy $\Psi$ loops in the relevant one-point function:
\bea\label{stationaryconditionnexttominimal}
& & -m^{2}+\lambda v^{2}+\frac{\lambda}{16\pi^{2}}\left[4m^{2}-10\lambda v^{2}-m^{2}\log\!\left(\frac{-m^{2}+\lambda v^{2}}{\mu^{2}}\right)-3m^{2}\log\!\left(\frac{-m^{2}+3\lambda v^{2}}{\mu^{2}}\right)\right.\nn\\
& &\left.\hspace{5.5cm}+\lambda v^{2}\log\!\left(\frac{-m^{2}+\lambda v^{2}}{\mu^{2}}\right)+9\lambda v^{2}\log\!\left(\frac{-m^{2}+3\lambda v^{2}}{\mu^{2}}\right)\right]\nn\\
& & \hspace{2cm}+\frac{\kappa}{16\pi^{2}}\left[-M^{2}-\kappa v^{2}+(M^{2}+\kappa v^{2})\log\!\left(\frac{M^{2}+\kappa v^{2}}{\mu^{2}}\right)\right]=0\,.
\eea
\paragraph*{Second derivatives:}
Similarly, the sums of the infinite series in (\ref{der2deltaV}) read (as before, we display only the factors with non-zero traces):
\bea
S_{RR}&=&\left(\begin{array}{ccc}
72 \lambda^{2}v^{2} & 0 & 0 \\
0 & 8\lambda^{2}v^{2} & 0\\
0 & 0 & 8\kappa^{2}v^{2}
\end{array}\right)\,,\\
S_{II}&=&4 \lambda(m^{2}-2\lambda v^{2})\log\!\left(\frac{-m^{2}+\lambda v^{2}}{-m^{2}+3\lambda v^{2}}\right)\left(\begin{array}{ccc}
1 & 0 & 0\\
0 &1 & 0 \\
0 & 0 & 0
\end{array}\right)\,,\\
S_{\Psi\Psi}&=&4\kappa^{2}v^{2}\frac{(M^{2}+\kappa v^{2})+(-m^{2}+3\lambda v^{2})}{(M^{2}+\kappa v^{2})-(-m^{2}+3\lambda v^{2})}\log\!\left(\frac{M^{2}+\kappa v^{2}}{-m^{2}+3\lambda v^{2}}\right)\left(\begin{array}{ccc}
1 & 0 & 0\label{SGG2}\\
0 &0 & 0 \\
0 & 0 & 1
\end{array}\right)\,.
\eea
Note that the last expression is regular even for $M^{2}+\kappa v^{2}\to -m^{2}+3\lambda v^{2}$ so it is only the log in $S_{II}$ that one should be careful about.

With this information at hand, one can readily calculate the second derivatives of the effective potential; the only non-trivial entries turn out to be the diagonal ones:
\bea
\left\langle\frac{\partial^{2}V_{\rm eff}}{\partial \phi_{R}^{2}}\right\rangle
&=& -m^{2}+3\lambda v^{2}+\frac{\lambda}{16\pi^{2}}\left[4m^{2}-10\lambda v^{2}-m^{2}\log\!\left(\frac{-m^{2}+\lambda v^{2}}{\mu^{2}}\right)-3m^{2}\log\!\left(\frac{-m^{2}+3\lambda v^{2}}{\mu^{2}}\right)\right.\nn\\
& &\left.\hspace{5cm}+3\lambda v^{2}\log\!\left(\frac{-m^{2}+\lambda v^{2}}{\mu^{2}}\right)+27\lambda v^{2}\log\!\left(\frac{-m^{2}+3\lambda v^{2}}{\mu^{2}}\right)\right]\nn\\
& & \hspace{2.2cm}+\frac{\kappa}{16\pi^{2}}\left[-M^{2}-\kappa v^{2}+(M^{2}+3\kappa v^{2})\log\!\left(\frac{M^{2}+\kappa v^{2}}{\mu^{2}}\right)\right]\,,
\label{secondderivativesnexttominimal0}\eea
\bea
\left\langle\frac{\partial^{2}V_{\rm eff}}{\partial \phi_{I}^{2}}\right\rangle
&=&  -m^{2}+\lambda v^{2}+\frac{\lambda}{16\pi^{2}}\left[4m^{2}-10\lambda v^{2}-m^{2}\log\!\left(\frac{-m^{2}+\lambda v^{2}}{\mu^{2}}\right)-3m^{2}\log\!\left(\frac{-m^{2}+3\lambda v^{2}}{\mu^{2}}\right)\right.\nn\\
& &\left.\hspace{5.5cm}+\lambda v^{2}\log\!\left(\frac{-m^{2}+\lambda v^{2}}{\mu^{2}}\right)+9\lambda v^{2}\log\!\left(\frac{-m^{2}+3\lambda v^{2}}{\mu^{2}}\right)\right]\nn\\
& & \hspace{2cm}+\frac{\kappa}{16\pi^{2}}\left[-M^{2}-\kappa v^{2}+(M^{2}+\kappa v^{2})\log\!\left(\frac{M^{2}+\kappa v^{2}}{\mu^{2}}\right)\right]\,,
\label{secondderivativesnexttominimal1}
\eea
\bea
\left\langle\frac{\partial^{2}V_{\rm eff}}{\partial \Psi^{2}}\right\rangle
&=& M^{2}+\kappa v^{2}-\frac{3M^{2}\rho}{4\pi^{2}}
+\frac{\kappa}{8\pi^{2}}\left[m^{2}-2v^{2}(\lambda+\kappa+3\rho)\right]\nn\\
& & 
\hspace{2cm}+\frac{1}{4\pi^{2}}\frac{3(m^{2}+M^{2})\rho+v^{2}(\kappa^{2}+3\kappa\rho-9\lambda\rho)}{m^{2}+M^{2}+v^{2}(\kappa-3\lambda)}
(M^{2}+\kappa v^{2})\log\!\left(\frac{M^{2}+\kappa v^{2}}{\mu^{2}}\right)\nn\\
& & \hspace{2cm}-\frac{\kappa}{16\pi^{2}}
\frac{m^{2}+M^{2}-3v^{2}(\kappa+\lambda)}{m^{2}+M^{2}+v^{2}(\kappa-3\lambda)}
(-m^{2}+3\lambda v^{2})\log\!\left(\frac{-m^{2}+3\lambda v^{2}}{\mu^{2}}\right)\nn\\
& & \hspace{2cm}+\frac{\kappa}{16\pi^{2}}
(-m^{2}+\lambda v^{2})\log\!\left(\frac{-m^{2}+\lambda v^{2}}{\mu^{2}}\right)\,.\label{secondderivativesnexttominimal2}
\eea
Again, the $\phi_{I}^{2}$ part (\ref{secondderivativesnexttominimal1}) exactly vanishes in the vacuum (\ref{stationaryconditionnexttominimal}) and, thus, we consistently recover the desired Goldstone zero in the scalar spectrum. Next, it is clear that there is indeed no singularity due to the last term in formula (\ref{secondderivativesnexttominimal2}) in the formal $m^{2}\to \lambda v^{2}$ limit  - this is to be expected as there is no way to cure a would-be real IR singularity in the heavy sector by finite shifts in formula (\ref{polemass}) because, given the shape of the $\phi_{I}-\Psi$ interaction term in (\ref{VWithPsiInComponents}), the only graph including the IR-singular Goldstone propagator that can contribute to $\Sigma_{\Psi\Psi}$ is a momentum-independent tadpole which drops from $\Sigma_{\Psi\Psi}(p^{2})-\Sigma_{\Psi\Psi}(0)$. Hence, it is legitimate to input the stationarity condition~(\ref{stationaryconditionnexttominimal}) into the one-loop part of~(\ref{secondderivativesnexttominimal2}) in the self-consistent approximate form $m^{2}=\lambda v^{2}$.

\subsection{The one-loop spectrum\label{nonminimalmodelspectrum}}
Expanding further the formula~(\ref{secondderivativesnexttominimal2}) in powers of $v/M$ the mass of the heavy scalar is given by
\be
\left\langle\frac{\partial^{2}V_{\rm eff}}{\partial \Psi^{2}}\right\rangle
= M^{2}+\frac{3M^{2}\rho}{4\pi^{2}}
\left[\log\!\left(\frac{M^{2}}{\mu^{2}}\right)-1\right]+\ldots\,,
\ee
where the ellipsis contains terms proportional to $v^{2}$ and smaller which in the case of our interest (i.e., $v\ll M$) play no role here.

For the zero-momentum-squared Higgs boson mass one has\footnote{As before, one should use the full solution of the stationarity equation~(\ref{stationaryconditionnexttominimal}) here but it is too difficult to deal with; rather than that we merely subtract (\ref{stationaryconditionnexttominimal}) from (\ref{secondderivativesnexttominimal0}) which is good enough to fix at least the leading polynomial structure.}:
\be\label{Higgszeromomentummassnexttominimal}
\left\langle\frac{\partial^{2}V_{\rm eff}}{\partial \phi_{R}^{2}}\right\rangle
= 2\lambda v^{2}+\frac{\lambda^{2}v^{2}}{8\pi^{2}}\log\!\left(\frac{-m^{2}+\lambda v^{2}}{\mu^{2}}\right)+\frac{9\lambda^{2}v^{2}}{8\pi^{2}}\log\!\left(\frac{-m^{2}+3\lambda v^{2}}{\mu^{2}}\right)+\frac{\kappa^{2}v^{2}}{8\pi^{2}}\log\!\left(\frac{M^{2}+\kappa v^{2}}{\mu^{2}}\right)
\ee
that, after the shift to the pole, provides the physical Higgs mass. 

Remarkably enough, even here the terms proportional to $M^{2}$ (i.e., those in the last rows of eqs.~(\ref{stationaryconditionnexttominimal}) and ~(\ref{secondderivativesnexttominimal0})) \ch{disappear}. Unlike in the minimal model discussed in Section~\ref{oneloopHiggssimplest}, this cancellation is non-trivial here\footnote{This should be expected though because, basically, the diagrammatics governing the renormalization of the VEV is virtually the same as that governing the renormalization of the two-point function; for a recent detailed study of this interplay in the framework of the  sigma model see, e.g., \cite{Lynn:2011aa}.} as there are two different singlet mass parameters entering the stationarity condition. These, however, simultaneously cancel among the one-point and two-point Green's functions. Hence, one can conclude that {\it even with an extra heavy field at play the Higgs mass remains to be sensitive to the high-scale physics only through its VEV}, as it is with any other field whose mass is generated by the spontaneous gauge symmetry breaking.

Let us also make it clear that the shift from the zero-momentum-scheme mass to the pole mass does not regenerate any explicit polynomial $M^{2}$-dependence (or $M^{2}\times$ log type of terms) in formula (\ref{Higgszeromomentummassnexttominimal}): although there are potentially large tadpoles contributing to $\Sigma_{HH}$, these are innocent in the difference $\Sigma_{HH}(p^{2})-\Sigma_{HH}(0)$ because of their momentum independence; similarly, the finite shifts due to the heavy momentum-dependent ``blobs'' behave essentially like $v^{2}[\log\!(M^{2}+p^{2})-\log\! M^{2}]$ and, thus, entertain the standard decoupling behaviour.  

Last remark concerns the $\mu$-dependence of formula~(\ref{Higgszeromomentummassnexttominimal}). The coefficient of the last term therein can be again obtained from a simple RGE argument: The only difference between the (scalar part of the) beta function in the minimal setting discussed in Section~\ref{oneloopHiggssimplest} and in the current situation is an extra contribution from the graph of the type (\ref{betafunctiongraphs}) with $\Psi$ instead of $H$ and/or $G$ propagating in the loop. Since, at the level of combinatorics, it is equivalent to that of the Goldstone type in (\ref{betafunctiongraphs}) with $\lambda$ swapped for $\kappa$ the coefficient of the last term in formula~(\ref{Higgszeromomentummassnexttominimal}) must be, up to this trivial difference, identical to the one of the second term therein. 
\section{The abelian Higgs model with two extra singlet scalars\label{sect:AHMplustwoscalars}}
It is important to note that the UV robustness of the Higgs boson mass (claimed to be at the same level as that of, e.g., the gauge fields) is a special feature of  just the Higgs type of fields and there is no reason for other elementary scalars unrelated to the spontaneous gauge symmetry breaking to be ``protected'' in any way by the relevant VEV. Qualitatively, this is not difficult to understand;  roughly speaking, \ch{for these fields the classical argument sketched in Section~\ref{sect:intro} applies because their large singlet mass term dominates the shape of the scalar potential in the relevant direction and, hence, the extra condition (\ref{onepointfunction}) is trivially satisfied}.  

To make this very clear one can repeat yet again the analysis in Section~\ref{sect:AHMplusscalar} with a pair of extra scalars $\Psi_{1}$, $\Psi_{2}$ with masses $M_{1}$ and $M_{2}$ instead of just one $\Psi$. Since the calculation is a straightforward repetition of what was done before let us quote here only the main result:
\bea
\left\langle\frac{\partial^{2}V_{\rm eff}}{\partial \Psi_{1}^{2}}\right\rangle
&=& M_{1}^{2}+\frac{3M_{1}^{2}\rho_{1}}{4\pi^{2}}
\left[\log\!\left(\frac{M_{1}^{2}}{\mu^{2}}\right)-1\right]+\frac{M_{2}^{2}\eta}{8\pi^{2}}
\left[\log\!\left(\frac{M_{2}^{2}}{\mu^{2}}\right)-1\right]\ldots\,,\nn\\
\left\langle\frac{\partial^{2}V_{\rm eff}}{\partial \Psi_{2}^{2}}\right\rangle
&=& M_{2}^{2}+\frac{3M_{2}^{2}\rho_{2}}{4\pi^{2}}
\left[\log\!\left(\frac{M_{2}^{2}}{\mu^{2}}\right)-1\right]+\frac{M_{1}^{2}\eta}{8\pi^{2}}
\left[\log\!\left(\frac{M_{1}^{2}}{\mu^{2}}\right)-1\right]\ldots\,.
\eea
In both cases, the ellipsis contains terms proportional to $v^{2}$ (or smaller) and $\eta$ is the coupling connecting the two scalars, i.e., ${\cal L}\ni \eta\, \Psi_{1}^{2}\Psi_{2}^{2}$. Obviously, unlike for the Higgs boson whose mass is (for a fixed VEV) only logarithmically $M$-dependent, in this case the heavier scalar (\ch{whichever of the two it is}) pushes up the mass of the lighter one by means of an explicit polynomial contribution. 

\section{Conclusions and comments}

In these notes we attempted to comment on some of the subtleties in calculating the scalar spectra of spontaneously broken gauge theories which, though often unnecessary to deal with in the order-of-magnitude types of estimates, have to be all taken into account if fully consistent results are sought for. In this respect, the effective potential approach can be seen as a great bookkeeping tool which naturally accounts for all the fine interplay among the renormalization effects in the propagator and in the vacuum of the theory. 

Sticking to the abelian Higgs model as a minimal setting \ch{encompassing} most of the salient features of a perturbative spontaneous gauge symmetry breakdown, we have exemplified in detail, for instance, how to handle the infinite series of nested commutators of derivatives of  field-dependent mass matrices emerging in the second derivatives of the effective potential, how to cope with the spurious IR divergences that may emerge in the self-consistent next-to-leading-order calculation of the Higgs mass, how does the shape of the derivatives of the effective potential relate to the $\beta$ and $\gamma$ functions of the theory, how infinitely more powerful are the effective potential methods in comparison to the modest purely diagrammatic approach in more complicated cases etc.

As a bonus, we have got a (hopefully) clear picture about the merit of the common lore that the elementary scalar masses are quadratically sensitive to the cut-off (representing, e.g., a higher energy scale where new degrees of freedom are integrated into the theory). As we saw, there is a clear difference between the physical mass formulae for the Higgs boson(s) and those for other generic scalars; as for the former, the large scale enters solely through the relevant VEV(s) (and, in this sense, the mass of the SM Higgs is as UV robust as that of any other SM field) while, for the latter, an explicit quadratic dependence on the large scale pops up. We put this into contrast with the usual wording of the hierarchy argument in support of many popular extension of the SM which  notoriously sticks to just the discussion of the explicit quadratic cut-off dependence of the scalar two-point function and almost never touches upon the other important ingredient at play, namely, the fate of the tree-level vacuum.

In view of this, there are several disturbing questions that one can not refrain from asking; for instance: Why do we actually demand  order-by-order perturbative stability of the unphysical parameters underpinning our calculations? Yes, we may need to adjust the {\it underlying} parameters to a high precision at the desired level in the perturbation theory in order to accommodate the data that we choose as our inputs, but why do we expect that such {\it not-directly-measurable} quantities would fall anywhere near (in whatever sense) those determined at a lower order of the perturbation theory? Alternatively, on a very practical level, why don't we care much about the cancellation of ``infinities'' in the loops contributing to the VEV shifts but, at the same time, we are so much concerned about the sizes of the finite remnants? Isn't this point of view just  overemphasised due to our inability to solve the theory?    

\ch{Traditionally, these concerns are addressed by working in frameworks where such questions do not even need to be asked, either because there is no high scale within the reach of the classical perturbative techniques (e.g., when there is no new physics assumed up to the Planck scale), or due to the emergence of new degrees of freedom acting as natural regulators of the wild perturbation theory behaviour (as it happens, for instance,  in the low-scale supersymmetry).}

\ch{Although this viewpoint is widely popular and very fruitful in practice it is not the only logically consistent option to deal with the hierarchy issue.} One could speculate that, maybe, we are just asking too much from physics which -- as a merely descriptive discipline --  should be expected to provide pre- or at least post-dictions of the outcome of {\it measurements} rather than far-reaching philosophical insights (which, however, we can not prevent ourselves from making up -- it's just too deep in our roots). If, for instance, all that we asked for was whether {\it correlations} between measurable quantities are stable order by order in the perturbation theory (for instance, the relations between the Higgs mass, the mass of the $Z$ boson, the the muon decay width and/or the ``strength'' of the neutral current interactions), the entire hierarchy problem would be gone because then the fate of the bare Higgs mass parameter underpinning the electroweak VEV becomes \ch{{\it physically irrelevant}.
}

However, this all would already bring us to the blurry region between physics and philosophy which the author doesn't feel qualified to enter.

\section*{Acknowledgments}
The work was partially supported by the Marie Curie Intra European Fellowship
within the 7th European Community Framework Programme
FP7-PEOPLE-2009-IEF, contract number PIEF-GA-2009-253119, by the Marie Curie Career Integration Grant within the 7th European Community Framework Programme
FP7-PEOPLE-2011-CIG, contract number PCIG10-GA-2011-303565, by the EU
Network grant UNILHC PITN-GA-2009-237920, by the Spanish MICINN
grants FPA2008-00319/FPA and MULTIDARK CAD2009-00064
(Consolider-Ingenio 2010 Programme), by the Generalitat
Valenciana grant Prometeo/2009/091 and by the Research proposal MSM0021620859 of the Ministry of Education, Youth and Sports of the Czech Republic.
I'm indebted to Luca di Luzio for reading through an early version of the manuscript.
\section*{Appendices}
\appendix
\section{Handling the infinite series of nested commutators\label{sect:handlingtheseries}}
Denoting for simplicity ${M}_{S}^{2}\equiv A$ and ${\partial {M}_{S}^{2}}/{\partial\phi_{a,b}}\equiv A_{a,b}$ the infinite series in equation~(\ref{der2deltaV}) reads
\be\label{series}
S_{ab}\equiv \sum_{m=1}^{\infty}(-1)^{m+1}\frac{1}{m}\sum_{k=1}^{m}{m\choose k}\left\{ A, A_{a}\right\}  
\left[A,..\left[A, A_{b}\right]..\right]\left(A-1\right)^{m-k}\,,
\ee
where the commutator in the $k$-th term is taken $k-1$ times. Interestingly, it is often the case that the $\left\{ A, A_{a}\right\} \left[A,..\left[A, A_{b}\right]..\right]$ part of the $k$-th term above (that we shall denote $f^{k}_{ab}$) can be written as a $(k-1)$-th power of a certain matrix $B$ commuting with $A$ which is further multiplied from the left by a constant matrix pre-factor $C$, i.e., 
\be\label{simplest}
f^{k}_{ab}=C\,B^{k-1},\qquad [A,B]=0\,.
\ee
Then, however, one can sum up the series~(\ref{series}) quite easily:
\bea\label{series2}
S_{ab}&=&C\sum_{m=1}^{\infty}(-1)^{m+1}\frac{1}{m}\sum_{k=1}^{m}{m\choose k}B^{k-1}\left(A-1\right)^{m-k}\nn\\
&=&CB^{-1}\sum_{m=1}^{\infty}(-1)^{m+1}\frac{1}{m}\sum_{k=1}^{m}{m\choose k}B^{k}\left(A-1\right)^{m-k}\nn\\
&=& CB^{-1}\sum_{m=1}^{\infty}(-1)^{m+1}\frac{1}{m}\left[\sum_{k=0}^{m}{m\choose k}B^{k}\left(A-1\right)^{m-k}-\left(A-1\right)^{m}\right]
\nn\\
&=&
CB^{-1}\left[\log\!(A+B)-\log\! A\right]\,. \label{derivation}
\eea
In the simplest case above, i.e., whenever one can implement~(\ref{simplest}), it is clear that $C=\left\{ A, A_{a}\right\}A_{b}$ and $B=A_{b}^{-1}[A,A_{b}]$. Note that in the derivation (\ref{derivation}) we also assumed that $B$ was invertible. Actually, the invertibility of $B$ is not really necessary as, in practice, the RHS of formula (\ref{series2}) can be defined by its limit even for a singular $B$. Finally, for $B=0$, the inner series in Eq.~(\ref{der2deltaV}) reduces to just its first term and the formal limit $\lim_{B\to 0}CB^{-1}\left[\log\!(A+B)-\log\! A\right]= CA^{-1}$ is also retained.
\section{The self-energies\label{sect:selfenergies} and the fate of the spurious IR divergences}
In what follows we shall use the asymmetric-phase\footnote{Recall that here we aim at the spurious IR divergence in formula in which the stationarity condition was also yet to be implemented -- working in this regime makes it clear that in both cases the singularities are of the same origin.}  lagrangian (\ref{treelagrangianasymmetric})
to calculate the scalar-sector contribution to the momentum-dependent part of the Higgs self-energy (which is all we need for the difference $\Sigma_{HH}(p^{2})-\Sigma_{HH}(0)$). The relevant Feynman diagrams are 
\be\label{selfenergygraphs}
-i\,\Sigma_{HH}(p^{2})=
\parbox{4cm}{\includegraphics[width=3.8cm]{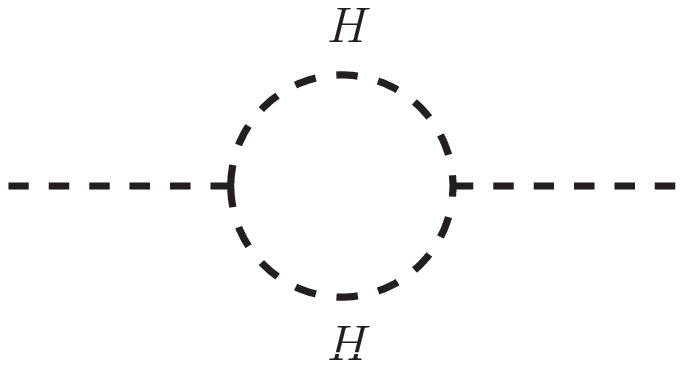}}+
\parbox{4cm}{\includegraphics[width=3.8cm]{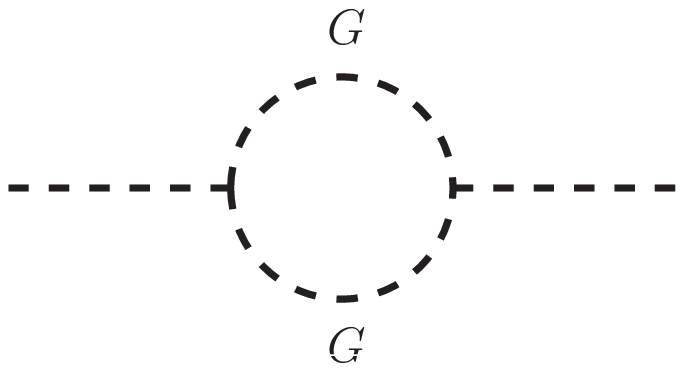}}+\ldots
\ee
and it is clear that only the second graph develops a spurious IR divergence in the $p^{2}\to 0$, $m^{2}\to \lambda v^{2}$ regime. 
The result of a simple calculation reads:
\be\label{SigmaExplicit}
\Sigma_{HH}(p^{2})=2\lambda^{2} v^{2}\left[9 I(p^{2},-m^{2}+3\lambda v^{2},-m^{2}+3\lambda v^{2})+I(p^{2},-m^{2}+\lambda v^{2},-m^{2}+\lambda v^{2})\right]\,,
\ee
where $I(p^{2},m_{1}^{2},m_{2}^{2})$ denotes the basic loop integral 
\be
\label{tosedivim}
I(p^{2},m_{1}^{2},m_{2}^{2})=
\frac{i}{16\pi^2}\left[\div-\intx\log\left(\frac{m_{1}^{2}x+m_{2}^{2}(1-x)-p^{2}x(1-x)}{\mu^{2}}\right) \right].
\ee
and $\div$ denotes the standard UV-divergent $\overline{\rm MS}$ structure in the dimensional regularization, i.e., $\div = \frac{1}{\varepsilon}-\gamma_{E}+\log 4\pi$ in $d=4-2\varepsilon$.
Hence, there is a term of the form $-\lambda^{2}v^{2}/8\pi^{2}\log[(-m^{2}+\lambda v^{2})/\mu^{2}]$ in the finite shift $\Sigma_{HH}(p^{2})-\Sigma_{HH}(0)$ which compensates the spurious IR divergence in the zero-momentum-squared formula (\ref{Higgszeromomentummass}) when the Higgs pole mass  is determined from Eq.~(\ref{polemass}).

\section{The Higgs sector beta and gamma functions\label{sect:betaandgamma}}
In this appendix we shall focus namely on the determination of the $\lambda$-dependent parts of the anomalous dimension $\gamma_{H}^{(\lambda)}$ of the Higgs field  and of its quartic coupling beta function $\beta_{\lambda}^{(\lambda)}$ in the minimal abelian Higgs model discussed in Section~\ref{oneloopHiggssimplest}; these are the key ingredients to formula~(\ref{treelevelshift})\footnote{\ch{It is not difficult to see that it  makes no difference whether these quantities are calculated in the perturbation theory corresponding to the symmetric- or asymmetric-phase lagrangian; for sake of simplicity we use the latter formalism but we do not explicitly draw the VEV-insertions in the propagators in the relevant Feynman graphs.}}.
\subsection{The $\lambda$-dependent part of the Higgs anomalous dimension\label{sect:anomalousdimension}}
It is quite simple to see that there is no contribution to $\gamma_{H}^{(\lambda)}$ at the one-loop level. The reason is that the trilinear couplings in the relevant scalar ``blob'' diagrams (\ref{selfenergygraphs}) 
are not dimensionless and, thus, the loop integration does not generate a momentum-squared-dependent contribution to the UV divergence. As a consequence, $\delta Z_{H}$ does not contain a term proportional to $\lambda$ and, thus, at one loop, $\gamma_{H}$ in a generic $R_{\xi}$ gauge receives only contributions from the gauge(+Goldstone) sector so one concludes 
$
\gamma_{H}^{(\lambda)}=0.
$
\subsection{The $\lambda$-dependent part of the Higgs quartic coupling beta function\label{sect:beta}}
For what follows it is convenient to write down the scalar part of the relevant lagrangian in the broken phase including counterterms
\be
{\cal L}_{H}\ni
\frac{1}{2}\partial_\mu H \partial^\mu H+ \delta Z_H\frac{1}{2}\partial_\mu H \partial^\mu H-\frac{1}{2}(2\lambda v^{2})H^{2}-\lambda vH^{3}-\lambda v H G^{2}-\frac{1}{4}{\lambda} H^4-\frac{1}{2}{\lambda}{G}^2H^2-
\frac{1}{4}\delta \lambda H^4\,,
\label{higgslagrangian}
\ee
where, for simplicity, we imposed the tree-level stationarity condition (which makes no difference here -- we will be anyway interested only in the UV divergences).
Given $\delta Z_{H}$ and $\delta\lambda$ the one-loop Higgs quartic coupling beta function can be written as
\be
\beta_{\lambda}=
-\lambda\frac{\partial K_\lambda}{\partial \log\!\mu}+
2\lambda\frac{\partial K_H}{\partial \log\!\mu}\,,
\ee
where $K_{\lambda}=\lambda^{-1}\delta\lambda$ and $K_{H}=\delta Z_{H}$. 
Note that in dimensional regularization one has  
\be
\frac{\partial K_{\lambda,H}}{\partial\log\!\mu}=
-2\vare K_{\lambda,H} + {\rm higher}\,\,{\rm order}\,\,{\rm terms}\,,
\ee
so, indeed, it is sufficient to consider the UV pole structure of the relevant diagrams.

Since $K_{H}$ at the one-loop level does not develop a purely $\lambda$-proportional UV divergent term (see Section~\ref{sect:anomalousdimension}) for our purposes here it is sufficient to evaluate only the proper vertex renormalization factor $K_{\lambda}$. 
This is done by considering the diagrams of the type:
\be\label{betafunctiongraphs}
\parbox{3cm}{\includegraphics[width=3cm]{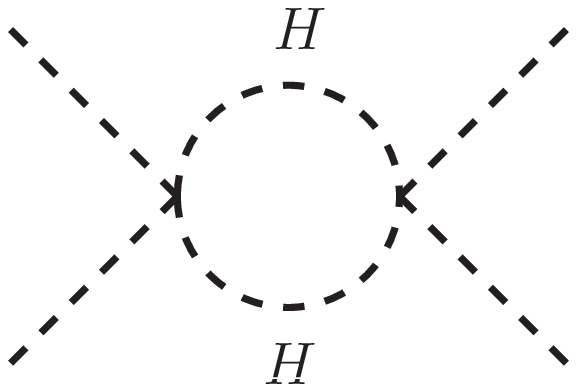}}+
\parbox{3cm}{\includegraphics[width=3cm]{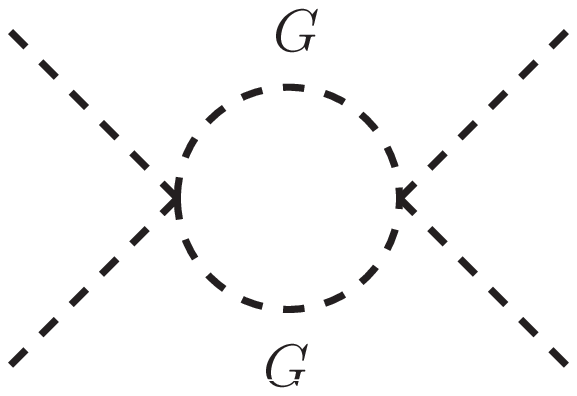}}
\ee
An elementary calculation yields the following contributions to the four-point function:
$$
\Delta \Gamma_{\lambda^2}^{(4)HH}=
{{\lambda}}^2\frac{27 i}{8\pi^2\vare}
+{\rm finite\, terms},
\qquad
\Delta \Gamma_{\lambda^2}^{(4)GG}=
{{\lambda}}^2\frac{3 i}{8\pi^2\vare}
+{\rm finite\, terms},
$$
which are compensated by the counterterm 
$
\Delta \Gamma_{\lambda^2}^{CT}=
-6i\,{\delta\lambda}
$
if and only if 
$
K_{\lambda}={5}\lambda/{8\pi^{2}\vare}+{\rm UV\ regular\, terms}
$.
Thus, one can conclude that the $\lambda$-dependent part of the one-loop Higgs quartic coupling beta function in the minimal abelian Higgs model reads
\be
\beta_{\lambda}^{(\lambda)}=\frac{5}{4\pi^{2}}\lambda^{2}.
\ee
\ch{
\section{The the second derivative of the effective potential and the pole mass\label{AppFiniteShifts}}
For the sake of completeness, let us just briefly recapitulate the derivation of the central formula (\ref{polemass}) here; this material can be, of course, found in the existing literature, see, e.g., \cite{Quiros:1997vk,Quiros:1999jp} and references therein. 

The physical (pole) mass is in any renormalization scheme $S$ given by the root of the $S$-renormalized inverse propagator
\be
\Gamma_{S}^{(2)}(p^{2})\equiv p^{2}-\mu_{S}^{2}-\Sigma_{S}(p^{2})=0\,.
\ee
Using the fact that the second derivative of the (dimensionally regularized $\overline{\rm MS}$) effective potential~(\ref{DeltaV}) in the vacuum ${\cal M}^{2}$ corresponds to $\Gamma_{\overline{\rm MS}}^{(2)}(0)$ one has
\be
{\cal M}^{2}=-\mu_{\overline{\rm MS}}^{2}-\Sigma_{\overline{\rm MS}}(0)\,,
\ee
and, hence, the physical mass corresponds to the root of the equation
\be
p^{2}-{\cal M}^{2}-\Sigma_{\overline{\rm MS}}(p^{2})+\Sigma_{\overline{\rm MS}}(0)=0\,.
\ee
This, however, is nothing but Eq.~(\ref{polemass})  up to a trivial generalisation to matrices.
}

\end{document}